\documentclass[12pt]{article}

\usepackage{xcolor}
\usepackage{graphics}
\usepackage{graphicx}

\usepackage{amsmath,amssymb}
\usepackage{siunitx}
\usepackage{times}
\usepackage{scicite}

\usepackage{tabularx}
\usepackage[normalem]{ulem}
\usepackage[utf8x]{inputenc}
\usepackage[font=small]{caption}
\usepackage{float}

\newenvironment{sciabstract}{
\begin{quote} \bf}
{\end{quote}}

\title{Polarization and phase textures in lattice plasmon condensates}

\author
{Jani M. Taskinen, Pavel Kliuiev, Antti J. Moilanen, P\"{a}ivi T\"{o}rm\"{a}$^{\ast}$\\
\\

\normalsize{Department of Applied Physics, Aalto University School of Science,}\\
\normalsize{P.O. Box 15100, Aalto, FI-00076, Finland}\\
\\
\normalsize{$^\ast$To whom correspondence should be addressed; E-mail:  paivi.torma@aalto.fi. }
}

\date{}

\begin{document}

\baselineskip24pt
\linespread{1.3}

\maketitle

\clearpage

\begin{sciabstract}
Polarization textures of light may reflect fundamental phenomena such as topological defects, and can be utilized in engineering light beams. Three main routes are applied during their creation: spontaneous appearance in phase transitions, steering by an excitation beam, or structural engineering of the medium. We present an approach that uses all three in a platform offering advantages that are not simultaneously provided in any previous system: advanced structural engineering, strong-coupling condensate with effective photonic interactions, as well as room temperature and sub-picosecond operation. We demonstrate domain wall polarization textures in a plasmonic lattice Bose-Einstein condensate, by combining the dipole structure of the lattice with a non-trivial condensate phase revealed by phase retrieval. These results open new prospects for fundamental studies of non-equilibrium condensation and sources of polarization-structured beams.
\end{sciabstract}

\clearpage
\noindent

\noindent Phase transitions and spontaneous symmetry breaking are associated with topological defects, for example vortices with windings of the phase of a superfluid or superconductor. A vector field with a (pseudo)spin or polarization degree of freedom allows an even richer set of topological defects such as skyrmions, merons, half-vortices, nodal lines and magnetic monopoles. These have been observed in solid state~\cite{tokura_magnetic_2020},
liquid crystal~\cite{alexander_colloquium_2012},
ultracold gas~\cite{stamper-kurn_spinor_2013},
and liquid Helium systems~\cite{volovik_universe_nodate}, 
as well as in polarization textures of light emerging from polariton condensates~\cite{hivet_half-solitons_2012,Manni2013,dufferwiel_spin_2015,cilibrizzi2016,donati_twist_2016}.
Polarization textures are also widely applied in beam engineering, lasing and holography~\cite{zhen2014,bliokh_spinorbit_2015,genevet_holographic_2015,rosales-guzman_review_2018,Wang2020}.

Three main approaches are typically used when creating polarization textures of light: 1) spontaneous appearance in a phase transition or a quench~\cite{Manni2013,cilibrizzi2016},
2) imposing the texture via an excitation or pump beam~\cite{donati_twist_2016} and
3) advanced structural engineering of the medium supporting the optical modes~\cite{bliokh_spinorbit_2015,sala_spin-orbit_2015,Wang2020}.
Polariton condensates in semiconductor systems are amenable for the first two, however, polarization textures have been observed only at cryogenic temperatures,
and complex structural engineering (e.g.,~\cite{sala_spin-orbit_2015,klembt_exciton-polariton_2018,su_observation_2020})
is technically demanding.
The latter is well developed in traditional photonic crystals and metamaterials, which, when combined with a gain medium, show lasing. However, strong-coupling condensation phenomena with related interactions have not been reported in those systems.
Here, we introduce a novel way of creating polarization textures that combines all three approaches in a platform that avoids the limitations of previously utilized systems. It is based on the design of optical dipoles in a plasmonic nanoparticle array, combined with a non-trivial phase structure of a Bose-Einstein condensate hosted by the lattice. Different polarization textures are switchable by the polarization of the beam pumping the condensate.
We experimentally demonstrate this new paradigm in a simple geometry and show that it leads to domain wall formation. For the first time for any kind of condensate, we reconstruct the BEC phase by a phase retrieval algorithm, avoiding interference measurements. Our system provides an extremely easy and versatile engineering of the geometry and unit cell of the lattice~\cite{hakala_lasing_2017,wang_rich_2018,kravets_plasmonic_2018,guo_lasing_2019}, along with
room temperature strong coupling condensation leading to effective interactions~\cite{de_giorgi_interaction_2018,Vakevainen2020}, and ultrafast sub-picosecond operation~\cite{hakala_bose-einstein_2018,Vakevainen2020}.  
The demonstrated new approach to polarization texture creation combines these assets in an unprecedented way, and is expected to be fruitful both in fundamental studies of non-equilibrium condensation phenomena~\cite{keeling_superfluidity_2017}
and in beam polarization engineering~\cite{bliokh_spinorbit_2015,genevet_holographic_2015,rosales-guzman_review_2018},
particularly when compact and ultrafast components are desired.

\begin{figure}
\centering
\includegraphics[width=0.7\textwidth]{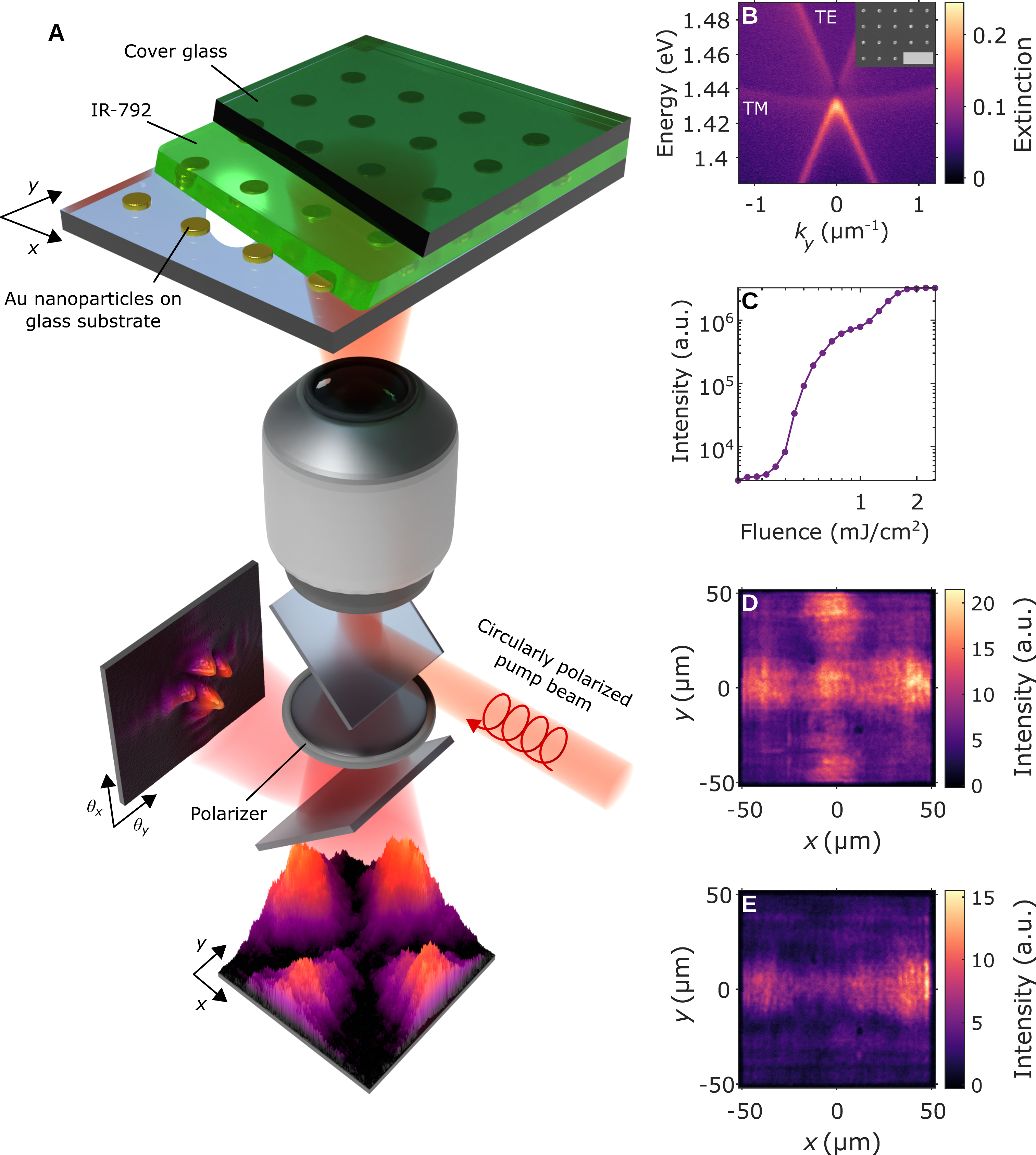}
\caption{\textbf{Spatially non-uniform condensation of lattice plasmon excitations.} (\textbf{A}) Illustration of the sample structure and the experimental configuration. Both the pumping of the molecules (IR-792) on the sample and collection of the sample photoluminescence are done via the same objective\cite{supplementary}, see Fig.~S1 for further details. (\textbf{B}) Extinction of a bare nanoparticle array immersed in index-matching oil measured as $(1 - T)$, where $T$ is the transmission. Here $k_y$ corresponds to the wave vector component parallel to the sample surface,
and is related to the emission angle $\theta_y$ and wavelength $\lambda_0$ as $k_y = 2 \pi / \lambda_0 \sin{\theta_y}$.
The dispersion of the modes shows the linear TE and parabolic TM surface lattices resonance modes crossing at the $\Gamma$-point. The inset is a scanning electron micrograph of the gold nanoparticles on a glass substrate, the scale bar is \SI{1}{\micro \meter}. (\textbf{C}) Total measured luminescence intensity of the sample as a function of pump fluence.
(\textbf{D}) Unpolarized and (\textbf{E}) horizontally polarized real-space images of the sample at 1.8~mJ/cm$^2$ pump fluence.
}
\label{fig:figure_1_system}
\end{figure}

An illustration of the system and the experiments is shown in Fig.~\ref{fig:figure_1_system}~(\textbf{A}).
A square array of golden nanoparticles fabricated on a glass substrate is immersed in a fluid and sealed with a cover glass. The fluid is either an index-matching oil for studies of the bare array, or a fluorescent dye solution such as a gain medium for the lasing and condensation measurements.
The bare arrays support collective plasmonic modes called surface lattice resonances (SLRs), which are hybrid modes consisting of the localized surface plasmon resonances of individual nanoparticles and the diffracted orders of the periodic lattice \cite{wang_rich_2018,kravets_plasmonic_2018}.
Excitations in the SLR modes are bosonic quasiparticles that have a mainly photonic nature, but also consist of collective electron oscillations in the nanoparticles; for their dispersion, see Fig.~\ref{fig:figure_1_system}~(\textbf{B}).
The $\Gamma$-point of the dispersion provides a band edge that may host lasing or condensation.

When the nanoparticle arrays are combined with emitters (e.g., dye molecules), the SLR modes remain intact (weak coupling) for low emitter concentrations while for high concentrations the strong coupling regime is reached and the excitations transfrom into polaritons, that is, hybrids of the SLR modes and emitter excitations~\cite{torma_strong_2015}. The emitters may be pumped externally and serve as a gain medium. Three modalities of coherent emission have been observed so far: 1) lasing at the weak coupling regime~\cite{hakala_lasing_2017,wang_structural_2018},
2) polariton lasing/condensation~\cite{de_giorgi_interaction_2018}, and
3) Bose-Einstein condensation (BEC) both at weak~\cite{hakala_bose-einstein_2018} and strong coupling~\cite{Vakevainen2020}. The BEC requires thermalization consisting of multiple molecule-light absorption and emission processes associated with loss of energy to molecular vibrational degrees of freedom~\cite{klaers_bose-einstein_2010}. The luminescence spectrum from the condensate shows a Bose-Einstein distribution, although the phenomenon is different from equilibrium BEC, as it occurs on the sub-picosecond scale. While output polarizations have been measured for previous nanoparticle array lasers and condensates, polarization textures have not been observed. 

Here, we work with the strong coupling BEC as introduced in~\cite{Vakevainen2020}. 
To reveal the fundamental properties of the condensate polarization structure, we use a highly symmetric geometry: the size and period of the lattice is the same in the $x$- and $y$-directions, and the particles are cylinders fully symmetric in the sample plane.
We combine the array with a fluorescent solution of IR-792 at a concentration of 80~mM, which leads to strong coupling. 
The sample is pumped at 800~nm using left circularly polarized laser pulses generated with an ultrafast Ti:sapphire laser.
The light radiated from the sample inherits the properties of the plasmonic excitations, therefore the condensate can be characterized via real-space, spectral, $k$-space and polarization-selective imaging. 
All the real-space data shown here are luminescence collected after a single pump pulse, that is, they correspond to single shot realizations of the condensate\cite{supplementary}.

Fig.~\ref{fig:figure_1_system}~(\textbf{C}) shows a distinct double-threshold behaviour in the total measured luminescence as a function of the pump fluence. The first threshold corresponds to polariton lasing, and the second one to strong coupling BEC~\cite{Vakevainen2020}; we focus here on the latter regime. The real-space intensity profile of the condensate including all polarizations is non-uniform and $x-y$ symmetric, see Fig.~\ref{fig:figure_1_system}~(\textbf{D}). With a linear polarizer in the $x$-direction, we obtain Fig.~\ref{fig:figure_1_system}~(\textbf{E}). When thermalizing towards the ground state ($\mathbf{k}=0$ band edge), the SLR excitations propagate due to their finite momentum $\mathbf{k}$. This together with the finite size of the array leads to a non-uniform condensate profile, which in~\cite{Vakevainen2020} was observed only in one direction (similar to Fig.~\ref{fig:figure_1_system}~(\textbf{E})) due to the use of a linearly polarized pump, which triggered the propagation only along one direction. Here, the circularly polarized pump leads to propagation and non-uniform condensate density in both $x$- and $y$-directions.

\begin{figure}
\centering
\includegraphics[width=\textwidth]{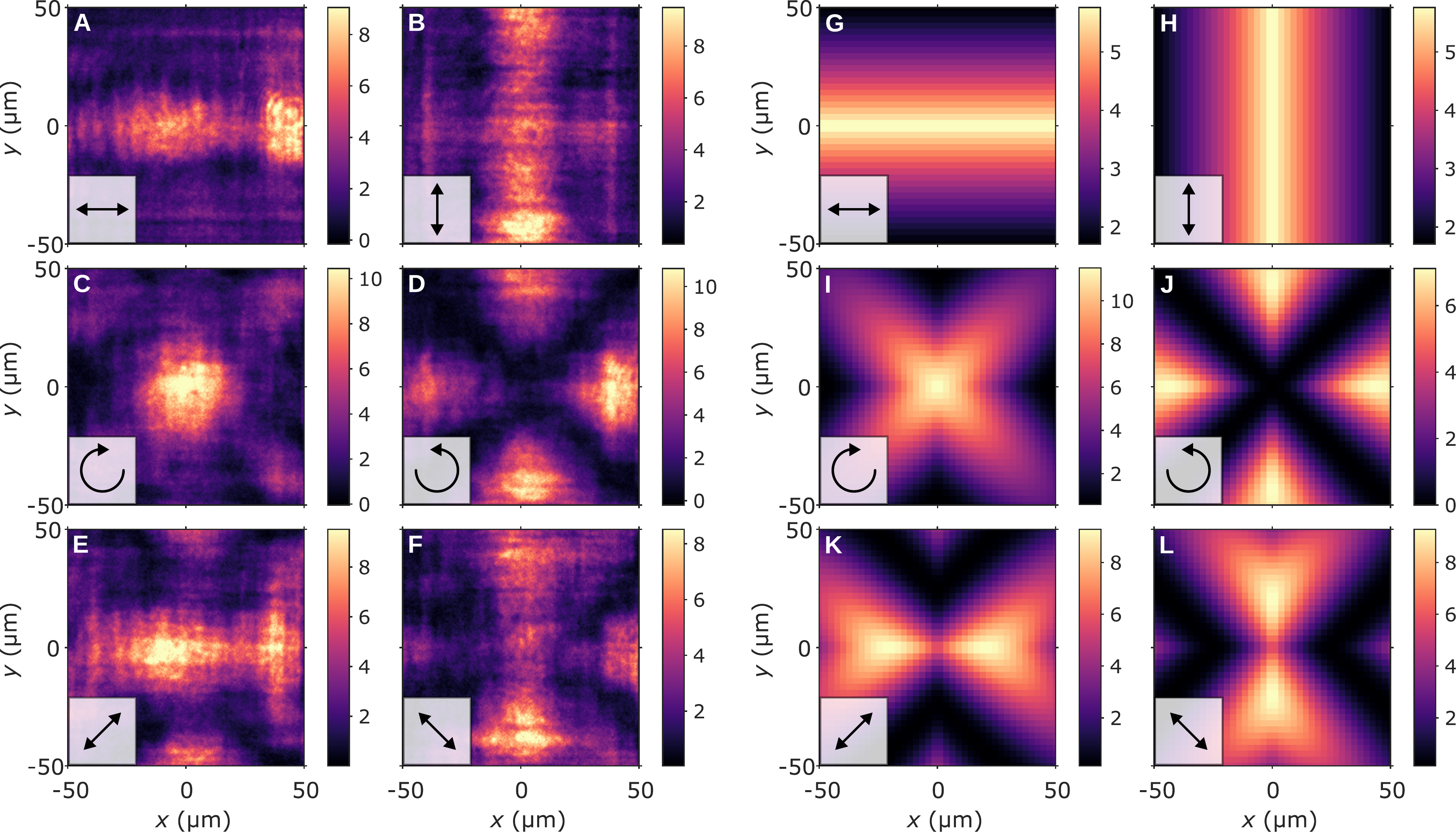}
\caption{\textbf{Real-space polarization patterns in a plasmonic condensate.} (\textbf{A}-\textbf{F}) Sample emission intensities under left circularly polarized pumping with the fluorescence imaged through (\textbf{A}) horizontal, (\textbf{B}) vertical, (\textbf{C}) right circular, (\textbf{D}) left circular, (\textbf{E}) diagonal and (\textbf{F}) antidiagonal polarizers. These polarizers are illustrated with black arrows. (\textbf{G}-\textbf{L}) Electric field intensities obtained by a theoretical model with polarizations corresponding to those in figures (\textbf{A}-\textbf{F}).
}
\label{fig:figure_2_patterns}
\end{figure}

Figs.~\ref{fig:figure_2_patterns}~(\textbf{A}-\textbf{F}) show the real-space intensity images of the pumped sample filtered using different polarizers.
Figs.~\ref{fig:figure_2_patterns}~(\textbf{A}-\textbf{B}) and (\textbf{E}-\textbf{F}) reflect the underlying $x-y$ symmetry of the system.
Remarkably, the right and left circularly polarized images (Figs.~\ref{fig:figure_2_patterns}~(\textbf{C}-\textbf{D})) show complementary intensity patterns, with right circularly polarized light being emitted from the centre and corners of the array, while luminescence close to the sides is mostly left circularly polarized. Figs.~\ref{fig:figure_2_patterns}~(\textbf{G}-\textbf{L}) are discussed later when we provide a theoretical model to explain the experimental findings. Note that right and left circular polarizations are superpositions of horizontal (here $x$) and vertical ($y$) linear polarizations $(|\updownarrow \rangle + i e^{i \varphi_{R/L}} |\leftrightarrow \rangle)/\sqrt{2}$, with the phase differing by $\pi$: $\varphi_R = \pi$, $\varphi_L = 0$. This, together with the observed change from right to left circular polarization over the array, would hint towards having a \textit{phase shift of the condensate} by $\pi$.
We have also used a diagonally polarized pump beam, which led to \textit{different} patterns (see Fig.~S2). This means that the patterns can be switched optically by femtosecond scale pulses.

\begin{figure}
\centering
\includegraphics[width=1\textwidth]{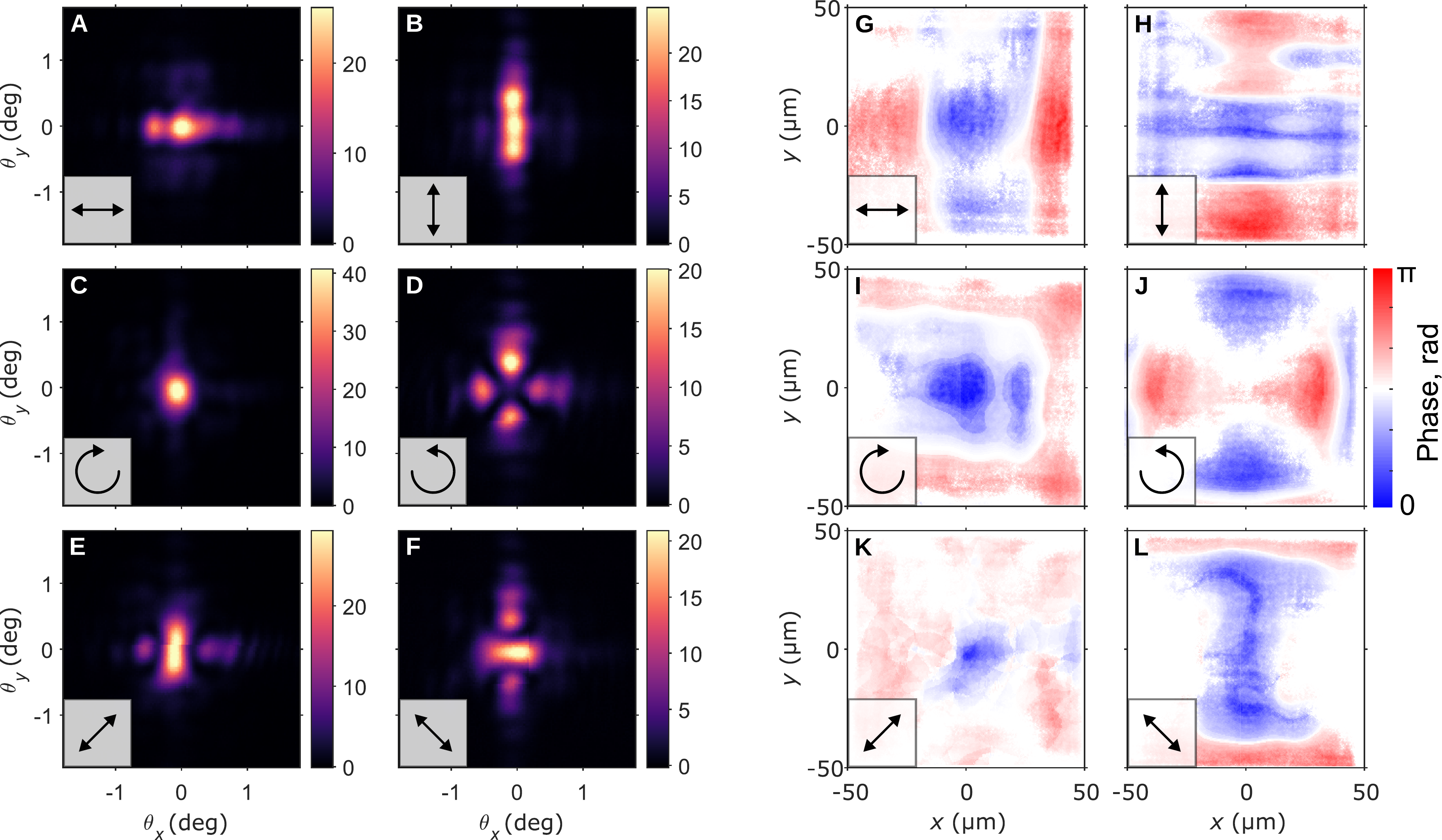}
\caption{\textbf{Reconstruction of phase in a plasmonic nanoparticle array.} (\textbf{A}-\textbf{F}) Two-dimensional $k$-space intensity images of the spatially non-uniform emission patterns shown in Figs.~\ref{fig:figure_2_patterns}~(\textbf{A}-\textbf{F}). Here $k_{x,y} = 2 \pi / \lambda_0 \sin{\theta_{x,y}}$. (\textbf{G}-\textbf{L}) Real-space phase distributions reconstructed from the real and $k$-space data using the Gerchberg-Saxton algorithm. The images display the phase differences in the sample with an arbitrary overall phase.
}
\label{fig:figure_3_kpatterns_and_phases}
\end{figure}

In addition to the real-space data, we capture
$k$-space ($k_x, k_y$) images of the luminescence (Figs.~\ref{fig:figure_3_kpatterns_and_phases}~(\textbf{A}-\textbf{F})) which display striking similarities to their real-space counterparts: Figs.~\ref{fig:figure_3_kpatterns_and_phases}~(\textbf{A}-\textbf{B}) and (\textbf{E}-\textbf{F}) reflect the $x-y$ symmetry, while the left and right circular polarization components (Figs.~\ref{fig:figure_3_kpatterns_and_phases}~(\textbf{C}-\textbf{D})) have momentum distributions distinct from each other, i.e., they cannot be made the same by a rotation. 
Together, the real-space and Fourier domain data sets (Figs.~\ref{fig:figure_2_patterns}~(\textbf{A}-\textbf{F}) and Figs.~\ref{fig:figure_3_kpatterns_and_phases}~(\textbf{A}-\textbf{F}), respectively) allow us to reconstruct the phase profiles of the differently polarized emission patterns by utilizing the Gerchberg-Saxton phase retrieval algorithm\cite{supplementary}. Figs.~\ref{fig:figure_3_kpatterns_and_phases}~(\textbf{G}-\textbf{L}) show the reconstructed real-space phase distributions.
Remarkably, the images display \textit{non-uniform phase profiles}: emission from the centre of the plasmonic array has obtained an opposite phase compared to the emission from the edges. Particularly interesting is the left circularly polarized case which shows the opposite phase appearing between adjacent edges. Starting from right circular polarization $(|\updownarrow \rangle - i |\leftrightarrow \rangle)/\sqrt{2}$ in the middle (Fig.~\ref{fig:figure_3_kpatterns_and_phases}~(\textbf{I})), then moving towards the edges in the $x$-direction and adding a $\pi$ phase shift to the $|\leftrightarrow\rangle$ component as indicated by Fig.~\ref{fig:figure_3_kpatterns_and_phases}~(\textbf{G}), one indeed obtains 
$(|\updownarrow \rangle + i |\leftrightarrow \rangle)/\sqrt{2}$, 
i.e.~left circular polarization. In contrast, moving in $y$ puts the minus sign in front of the $|\updownarrow\rangle$ component, leading to left circular polarization \textit{with an overall phase difference of $\pi$ compared to the other edge}, namely $-(|\updownarrow \rangle + i |\leftrightarrow \rangle)/\sqrt{2}$, in full agreement with Fig.~\ref{fig:figure_3_kpatterns_and_phases}~(\textbf{J}). This consistency between independently reconstructed images provides immediate proof of the robustness of the phase retrieval method.

\begin{figure}
\centering
\includegraphics[width=0.8\textwidth]{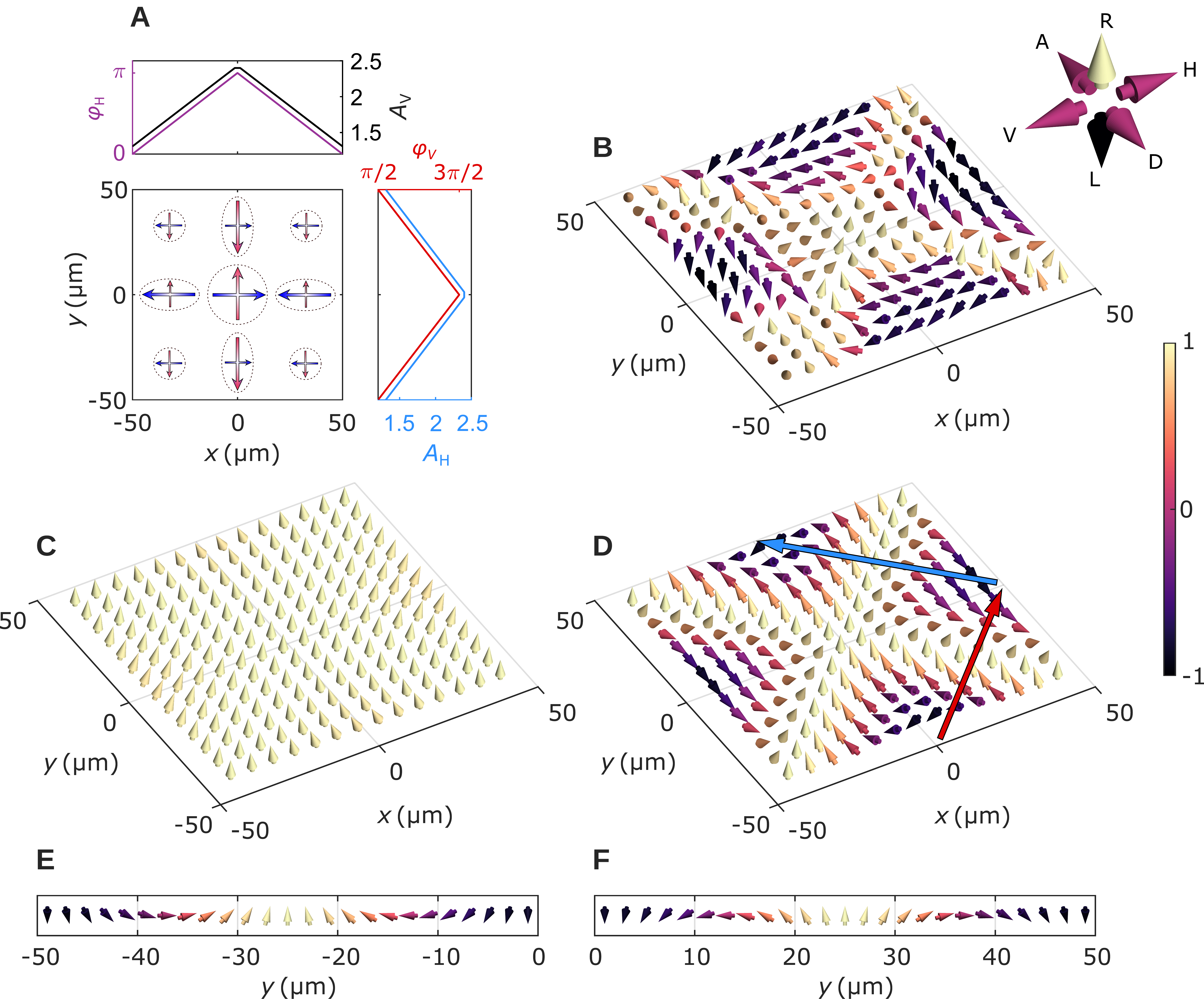}
\caption{\textbf{Jones vector grid and Stokes vector comparison.} (\textbf{A}) Schematic of the array model of Jones vectors depicting the polarization states of the plasmonic condensate. The arrows in different points of the array depict the amplitudes (arrow length) and phases (arrow direction) of the horizontal (blue arrows) and vertical (red arrows) polarization components. The amplitude profile used in both components (black and blue straight lines) increases towards the sample centre. The phase profiles are approximated linearly with a $\pi$-phase increase from the edges towards the array centre (purple and red lines). (\textbf{B}-\textbf{D}) Stokes vectors illustrating the polarization state at different points on the array, calculated from (\textbf{B}) the experimental real-space polarization patterns, and from theoretical dipole maps with (\textbf{C}) constant and (\textbf{D}) non-uniform phases. The relation between the polarization states and Stokes vectors is given in the arrow axes on the right. The length of each Stokes vector along the circular polarization axis is illustrated with a color scale, where -1 corresponds to left circular and +1 to right circular polarization. (\textbf{E}-\textbf{F}) Stokes vector winding along the (\textbf{E}) red and (\textbf{F}) blue arrows shown in (\textbf{D}). The vectors are viewed parallel to the array surface towards negative $x$-values.
}
\label{fig:figure_4_model}
\end{figure}

Based on the non-uniform phase distributions obtained from the phase retrieval algorithm, we construct a simple yet highly effective model of the polarization states of the lattice plasmons.
The array is modeled as a 2D grid of Jones vectors depicting the polarization of the emitted light. As a basis, we use the amplitudes $A_{\text{H}}$ and $A_{\text{V}}$, and phases $\varphi_{\text{H}}$ and $\varphi_{\text{V}}$, of the horizontally ($x$) and vertically ($y$) polarized electric field components.
In the collective SLR lattice modes, the magnitude of a nanoparticle dipole oscillating in the $y$-direction ($x$-direction) decreases when moving from the centre of the array towards the edges in the $x$-direction ($y$-direction). This is because the nanoparticles closer to the edges receive no radiation from outside the array. Fig.~\ref{fig:figure_4_model}~(\textbf{A}) shows a schematic of our model. Based on the simple argument of nanoparticles receiving no radiation beyond the array boundaries, the decrease is linear and of a factor of two. Alternatively, one can utilize the measured real-space intensity data in the model; this is discussed in the supplementary materials (see Fig.~S3).
Motivated by the results of the phase reconstruction, we apply a linear approximation for the phase components: $\varphi_{\text{H}}$ is varied from 0 to $\pi$ and back as a function of $x$ (purple line), and $\varphi_{\text{V}}$ from $\pi/2$ to $3\pi/2$ and back as a function of $y$ (red line). The relative phase difference of $\pi/2$ between $\varphi_{\text{H}}$ and $\varphi_{\text{V}}$ is chosen to correspond to right circular polarization at the centre of the array, as observed experimentally and anticipated from the pump polarization.

In order to compare the theoretical model to the measured real-space intensities, Jones calculus is applied to the 2D grid of Jones vectors, and the resulting electric field intensities of linear and circular polarization states are plotted in Figs.~\ref{fig:figure_2_patterns}~(\textbf{G}-\textbf{L}).
Remarkably, the combination of non-uniform amplitude and phase distributions allows the model to qualitatively reproduce the real-space polarization patterns observed in the plasmonic condensate.
Moreover, removing the $\pi/2$-phase difference between $\varphi_{\text{H}}$ and $\varphi_{\text{V}}$, which corresponds to diagonal polarization in the centre of the array, indeed leads to images that are consistent with the experimental intensity patterns observed with a diagonally polarized pump beam (see Fig.~S2).
This demonstrates that the observed phase distributions can be switched by the polarization state of the pump.

So far, we have investigated projections of the condensate emission on different polarization components. Now we also calculate the Stokes vectors $\mathbf{S} = (S_1, S_2, S_3)$, characterizing the pseudospin nature of polarization and how it evolves at different points on the array. The vector components are given by
\begin{equation}
    S_1 = \frac{I_{\text{H}} - I_{\text{V}}}{I_{\text{H}} + I_{\text{V}}}
    \text{,  }\quad 
    S_2 = \frac{I_{\text{R}} - I_{\text{L}}}{I_{\text{R}} + I_{\text{L}}}
    \text{,  }\quad 
    S_3 = \frac{I_{\text{D}} - I_{\text{A}}}{I_{\text{D}} + I_{\text{A}}} ,
\end{equation}
where $I_\sigma$ is the measured luminescence intensity and $\sigma$ corresponds to the six different polarization states (horizontal (H), vertical (V), right circular (R), left circular (L), diagonal (D), and antidiagonal (A)).
The pseudospin textures given by the experimental data, as well as those predicted by the theoretical model with uniform and non-uniform phase profiles, are shown in Figs.~\ref{fig:figure_4_model}~(\textbf{B}-\textbf{D}).
The spin texture plotted with a uniform phase distribution (Fig.~\ref{fig:figure_4_model}~(\textbf{C})) does not produce the complex spin textures calculated from the experimental results  (Fig.~\ref{fig:figure_4_model}~(\textbf{B})).
In contrast, there is a striking resemblance between these experimentally observed pseudospin patterns and the theoretical model with the \textit{non-uniform} phase distribution (Fig.~\ref{fig:figure_4_model}~(\textbf{D})), which demonstrates the importance of the discovered phase profiles in the system.

The observed pseudospin texture (Fig.~\ref{fig:figure_4_model}~(\textbf{B})) and the corresponding theory prediction \linebreak (Fig.~\ref{fig:figure_4_model}~(\textbf{D})) show clear domain walls separating four regions with mostly left circularly polarized emission. Fig.~\ref{fig:figure_4_model}~(\textbf{E}) shows the Stokes vector orientations following the red arrow across one of the domain walls in Fig.~\ref{fig:figure_4_model}~(\textbf{D}), and windings of $2\pi$ and $\sim1.6\pi$ are observed along the R-D and R-H planes, respectively. We do not observe a full rotation along the R-H plane as the amplitudes of the horizontally and vertically polarized components are different at the edges. These windings are reversed in the adjacent domain wall (blue arrow) shown in Fig.~\ref{fig:figure_4_model}~(\textbf{F}), and, following a closed loop around the centre of the sample, the total winding number becomes zero.
Here, in a system with simple square lattice geometry, the pseudospin texture is of non-topological nature.
Given the broad tunability of the plasmonic nanoparticle array and the dependence of the phase profiles on the pump polarization, the creation of topologically non-trivial textures is a feasible goal.

In summary, we have observed polarization textures arising from an interplay between a structured optical medium and a non-uniform Bose-Einstein condensate phase. One ingredient of the textures is the finite size of the periodic array, which causes the nanoparticle dipoles to weaken towards the edges. This alone, however, would lead to nothing but unremarkable effects on the polarization properties (c.f.~Fig.~\ref{fig:figure_4_model}~(\textbf{C})). For the observed prominent domain wall structures (Figs.~\ref{fig:figure_4_model}~(\textbf{E}-\textbf{F})), an additional element is crucial: the \textit{non-uniform phase} of the condensate. We revealed a zero-to-$\pi$ phase change between the central and edge parts of the array with a Gerchberg-Saxton algorithm. In addition to being essential for explaining the textures, this constitutes the first experimental determination of a condensate phase by computational imaging, proposed earlier by theory~\cite{meiser_2005,kosior_2014}.
This achievement puts forward an attractive alternative to measurements of phase by interference, replacing complex experiments by a robust computational approach.

For our proof-of-concept demonstration of polarization textures, we used a $C_2$ symmetric ($x-y$ symmetric) system. Future design possibilities include lattices with different geometry, size, and structure of the unit cell to realize new combinations of broken or competing symmetries, artificial gauge fields, and pseudospin-orbit coupling; different material choices such as dielectrics are also feasible~\cite{Heilmann2020,kuznetsov_optically_2016,staude_metamaterial-inspired_2017}.
Importantly, the simple theoretical framework introduced here allows fast and intuitive planning of the desired textures. The expected qualitative behaviour of the field intensities for different polarizations can be determined from the geometry of the lattice, and straightforwardly generalized to higher order multipolar nanoparticle modes and more complex unit cells. Such approach allows one to explore and plan polarization textures that various lattice configurations, together with different phase profiles, can produce.
To exploit the condensate phase as a design degree of freedom, further studies are needed to understand its formation: what are the roles of the finite lattice size, dispersion, electronic-vibrational-photonic strong coupling, and related non-linearities~\cite{arnardottir_multimode_2020}
in the apparently soliton-like phase and intensity profile of the condensate? Our results open new prospects for fundamental studies of vectorial (pseudospin) non-equilibrium condensates~\cite{keeling_superfluidity_2017} and topological photonics~\cite{ozawa_topological_2018},
as well as for tailoring bright coherent beams with complex polarization properties. The room temperature operation, straightforward sample fabrication, and ultrafast switching by the pump polarization are important assets. On-chip pumping would complete the list; combining plasmonic nanoparticle arrays with organic materials amenable to electrically induced gain~\cite{Daskalakis2019ACSPhoton} is obviously a worthwhile future research direction.

\pagebreak

\section*{Acknowledgments:}

\textbf{Funding:} This work was supported by Academy of Finland under project numbers 303351, 307419, 327293, 318987 (QuantERA project RouTe), 318937 (PROFI), and 320167 (Flagship Programme, Photonics Research and Innovation (PREIN)), and by the Centre for Quantum Engineering (CQE) at Aalto University. A.J.M. acknowledges financial support by the Jenny and Antti Wihuri Foundation.
Part of the research was performed at the OtaNano Nanofab cleanroom (Micronova Nanofabrication Centre), supported by Aalto University. Computing resources were provided by the Triton cluster at Aalto
University.
\textbf{Author contributions}: P.T. initiated and supervised the project. J.M.T. built the experiment,
fabricated the samples and analysed the data. J.M.T. and A.J.M. did the
measurements. P.K. performed the phase retrieval. J.M.T. and P.T. developed the
theoretical model to describe the experimental findings. All authors
discussed the results. P.T. and J.M.T. wrote the manuscript together with all
authors.  \textbf{Competing interests:} The authors declare no competing interests. \textbf{Data and materials availability}: All data used in this study are available at https://zenodo.org with the identifier DOI: 10.5281/zenodo.XXXXXXX.

\pagebreak
\section*{Supplementary Materials:}
Materials and Methods\\
Supplementary Text\\
Figures S1 to S3\\
References \textit{(38 - 42)}

\pagebreak

\renewcommand{\thefigure}{S\arabic{figure}}
\renewcommand{\thetable}{S\arabic{table}}
\renewcommand{\theequation}{S\arabic{equation}}
\setcounter{figure}{0}

\section*{Materials and methods}

\subsection*{Sample fabrication}

\nocite{tokura_magnetic_2020,alexander_colloquium_2012,stamper-kurn_spinor_2013,volovik_universe_nodate,hivet_half-solitons_2012,Manni2013,dufferwiel_spin_2015,cilibrizzi2016,donati_twist_2016,zhen2014,bliokh_spinorbit_2015,genevet_holographic_2015,rosales-guzman_review_2018,Wang2020,sala_spin-orbit_2015,klembt_exciton-polariton_2018,su_observation_2020,hakala_lasing_2017,wang_rich_2018,kravets_plasmonic_2018,guo_lasing_2019,de_giorgi_interaction_2018,Vakevainen2020,hakala_bose-einstein_2018,keeling_superfluidity_2017,torma_strong_2015,wang_structural_2018,klaers_bose-einstein_2010,supplementary,meiser_2005,kosior_2014,Heilmann2020,kuznetsov_optically_2016,staude_metamaterial-inspired_2017,arnardottir_multimode_2020,ozawa_topological_2018,Daskalakis2019ACSPhoton}

Square arrays of Au nanoparticles are fabricated on borosilicate glass slides using electron beam lithography. A polymethyl methacrylate (PMMA) layer, which is spin-coated and baked solid on the glass substrates, is covered with 10~nm of evaporated aluminum and patterned using an electron beam. The aluminum layer is then removed using 50\% AZ 351B developer and the PMMA layer is developed by immersion in 1:3 methyl isobutyl ketone:isopropanol solution. A thin (2~nm) titanium adhesion layer and a 50~nm gold layer are evaporated on the patterned slide, and excess PMMA and metal are removed by acetone lift-off. The array size is 100~$\times$~100~\SI{}{\micro \meter^2}, and the nominal periodicity 568~nm, which sets the $\Gamma$-point energy at $\sim1.44$~eV. The height and diameter of the cylindrical nanoparticles forming the array are 50~nm and 105~nm, respectively.

The samples are prepared for measurements by sealing the nanoparticles between the substrate and a cover glass slide using a circular silicone isolator, whose thickness is 0.8~mm. For transmission measurements, the isolator is filled with index-matching oil; for condensation measurements, it is filled with an 80~mM solution of IR-792 perchlorate dissolved in 1:2 dimethyl sulfoxide:benzyl alcohol mixture. The solvent has a matching refractive index with the glass slides ($n = 1.52$).

\subsection*{Experimental setup}

\begin{figure}[h]
\centering
\includegraphics[width=\textwidth]{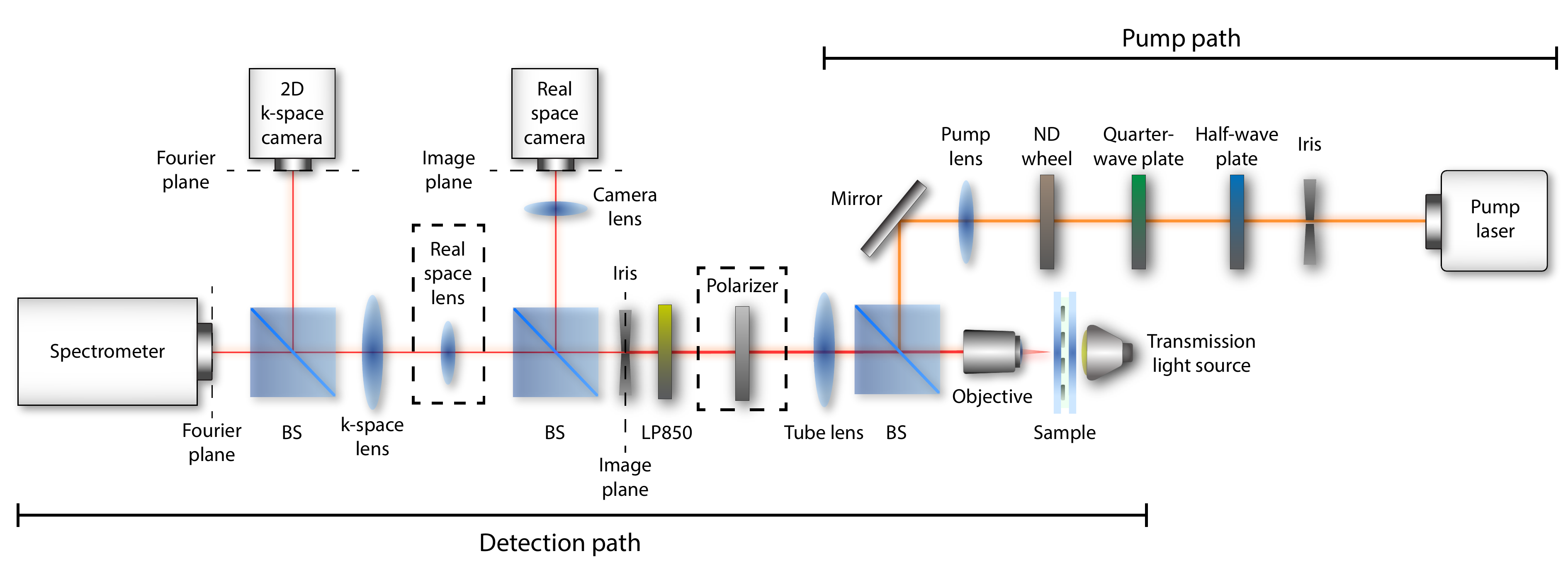}
\caption{Schematic of the experimental setup used in the angle- and energy-resolved intensity measurements. The spectrometer and two cameras allow the setup to simultaneously measure the spectral information of the emitted light and capture Fourier and real-space images of the sample. Optional components are marked with dashed rectangles. Here, BS stands for beamsplitter and ND for neutral density, and LP850 refers to a longpass filter with a cutoff wavelength of 850~nm.}
\label{fig:figure_S1_measurement_setup}
\end{figure}

A detailed schematic of the setup is shown in Fig.~\ref{fig:figure_S1_measurement_setup}. The measurement setup can be used to measure both angle-resolved $k$-space spectra as well as real-space spectra with minor modifications. Light exiting the sample is collected using an infinity corrected objective (10x, 0.3 NA) together with a compatible tube lens. An optional polarizer may be placed after the tube lens to limit the measurement to a single polarization state. Here, the polarization is defined from the point of view of the source. A long pass filter (cutoff wavelength at 850~nm) is used in the detection path to filter out pump reflections. Light from the sample is spatially restricted to the nanoparticle array using an adjustable iris in front of the cameras.

In $k$-space measurements, the back focal plane of the objective is focused to the entrance slit of a spectrometer such that each point on the slit corresponds to a specific emission angle $\theta_y$. The angle is related to momentum as $k_y = k_0\sin{\theta_y} = 2 \pi / \lambda_0 \sin{\theta_y}$, where $k_0$ and $\lambda_0$ are the free space wavenumber and wavelength, respectively. This allows the 2D charge-coupled device (CCD) array inside the spectrometer to measure a spectrum at different values of $k_y$ simultaneously. In addition, two fast complementary metal–oxide–semiconductor (CMOS) cameras are used to take direct real- and momentum-space images of the sample. The long pass filter limits the emission collected by the CMOS cameras to energies below 1.46~eV. In real-space measurements, an additional lens is placed after the tube lens, which causes the real-space image of the sample to be formed at the entrance slit of the spectrometer. In this case, each point on the slit corresponds to a specific $y$-coordinate of the sample.

In the condensation experiment, the sample is pumped optically using ultrafast laser pulses (50~fs pulse duration, 800~nm centre wavelength (1.55 eV energy)), that are left circularly polarized. However, as the pulses reflect off the sample, they are observed as right circularly polarized in the detection path. Since the repetition rate of our pump pulses is 1~kHz, setting the integration time of our CMOS cameras to 1~ms allows us to capture luminescence from a single realization of the condensate. An iris is used to spatially crop the pump beam, which is then focused on the nanoparticle array through the objective with the help of an additional pump lens. The polarization state of the pulses is controlled using motorized quarter- and half-wave plates on the pump path. Pump fluence is varied using a neutral density wheel. In transmission measurements, the array is illuminated using a broadband halogen light source.

\subsection*{Phase retrieval}

Prior to phase retrieval, real- and $k$-space images shown in Figs.~2~(\textbf{A}-\textbf{F}) and Figs.~3~(\textbf{A}-\textbf{F}) were pre-processed using standard procedures\cite{Kliuiev_2016}. Each dataset was centred in the computational domain.
Fourier data were centred by finding a local maximum or a local minimum in the vicinity of the physical centre of the $k$-space intensity distribution. The centre of the real-space data was found by applying a watershed segmentation algorithm to the real-space intensity image, and computing the centre of mass of the segmented region. Real-space data were re-sampled to fulfill the relationship between the pixel sizes in the object and Fourier domains as set by the digital Fourier transformation. The background noise (average 3100 counts in the object domain and 2200 counts in the $k$-space) was subtracted from each pixel; 3300 counts were subtracted from Fig.~2~(\textbf{C}) and 2150 counts were subtracted from Figs.~3~(\textbf{B}-\textbf{C}), as this led to a better convergence. 

The phase reconstruction was performed by the Gerchberg-Saxton phase retrieval algorithm \cite{G&S1972}.  The object-domain constraint was the square root of the processed real-space intensity distribution (Figs.~2~(\textbf{A}-\textbf{F})), and the Fourier constraint was the
square root of the processed $k$-space intensity distribution (Figs.~3~(\textbf{A}-\textbf{F})). The linear oversampling ratio was $\O\approx 11$ and thus fulfilled the oversampling condition \cite{Miao1998}. In total, we performed 1000 independent reconstruction rounds with different initial random phase distributions in the $k$-space. The random phases were generated from a uniform distribution between -$\pi$ and $\pi$. Each reconstruction round comprised 100 iterations, as this number of iterations was sufficient for the algorithm to converge. Only $5 \%$ of the reconstructed phase distributions having the lowest error metric in the $k$-space\cite{Fienup1978} were selected out of 1000 reconstructions and averaged following standard protocol\cite{Kliuiev2018}. For averaging purposes, the phase value in the centre of the computational domain was used as a reference, and accounted for an arbitrary global phase shift in the reconstructed images. The resulting reconstructed real-space phase distributions shown in Figs.~3~(\textbf{G}-\textbf{L}) were weighted with the corresponding amplitude values (square root of the processed Figs.~2~(\textbf{A}-\textbf{F})) for illustration purposes.

\section*{Supplementary Text}

\subsection*{Diagonal pumping}

\begin{figure}
\centering
\includegraphics[width=0.5\textwidth]{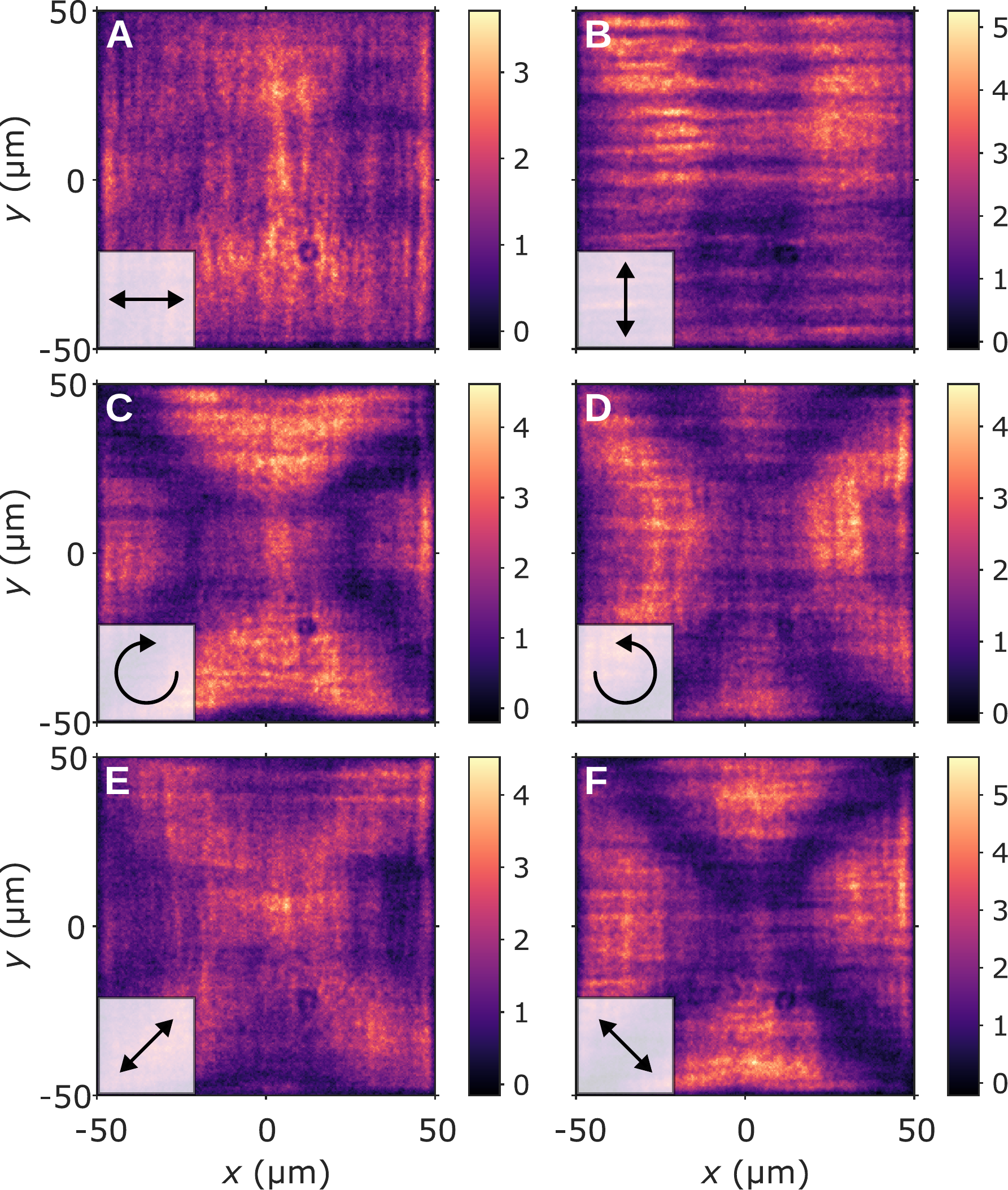}
\caption{\textbf{Polarization patterns under diagonal pumping.} (\textbf{A}-\textbf{F}) Real-space intensities emitted by the diagonally pumped sample measured through (\textbf{A}) horizontal, (\textbf{B}) vertical, (\textbf{C}) right circular, (\textbf{D}) left circular, (\textbf{E}) diagonal and (\textbf{F}) antidiagonal polarizers.}
\label{fig:figure_S2_diagonal_pump_RS_images}
\end{figure}

In addition to the circularly polarized pumping scheme described in the main text, we also investigate sample luminescence under diagonally polarized pumping. The resulting real-space images are presented in Fig.~\ref{fig:figure_S2_diagonal_pump_RS_images}, where the black arrows illustrate the polarization state of luminescence. The overall measured intensity is lower compared to circularly polarized pumping, while the vertically and horizontally polarized components do not show a clear accumulation of SLR excitations to the centre of the sample. A comparison between Figs.~2 and \ref{fig:figure_S2_diagonal_pump_RS_images} shows that, although both pump configurations show the same assortment of patterns, the textures associated with circular and diagonal polarization states are switched when we change the pump polarization: if patterns (\textbf{D}) and (\textbf{C}) are swapped with (\textbf{E}) and (\textbf{F}) in Fig.~2, we reach the arrangement of textures in Fig.~\ref{fig:figure_S2_diagonal_pump_RS_images}. The phase delay between the vertical and horizontal components of the pump is inherited to the plasmonic excitations and affects 
the observed polarization textures. 
A plausible explanation for this is the start of the lasing/condensation process by stimulated rather than spontaneous emission: The pump beam is not resonant with the plasmonic modes and thus mainly excites the molecules, however, it also drives off-resonant weak excitations in the nanoparticles. These excitations stimulate the initial molecular emission, and their polarization properties thus influence the lasing or condensate state. Initiation of the condensation process by stimulated rather than spontaneous emission is in accordance with the ultrafast timescales and stimulated nature of the thermalization that were observed in~\cite{Vakevainen2020}.

\subsection*{Theoretical model with experimental amplitude profiles}

\begin{figure}
\centering
\includegraphics[width=0.8\textwidth]{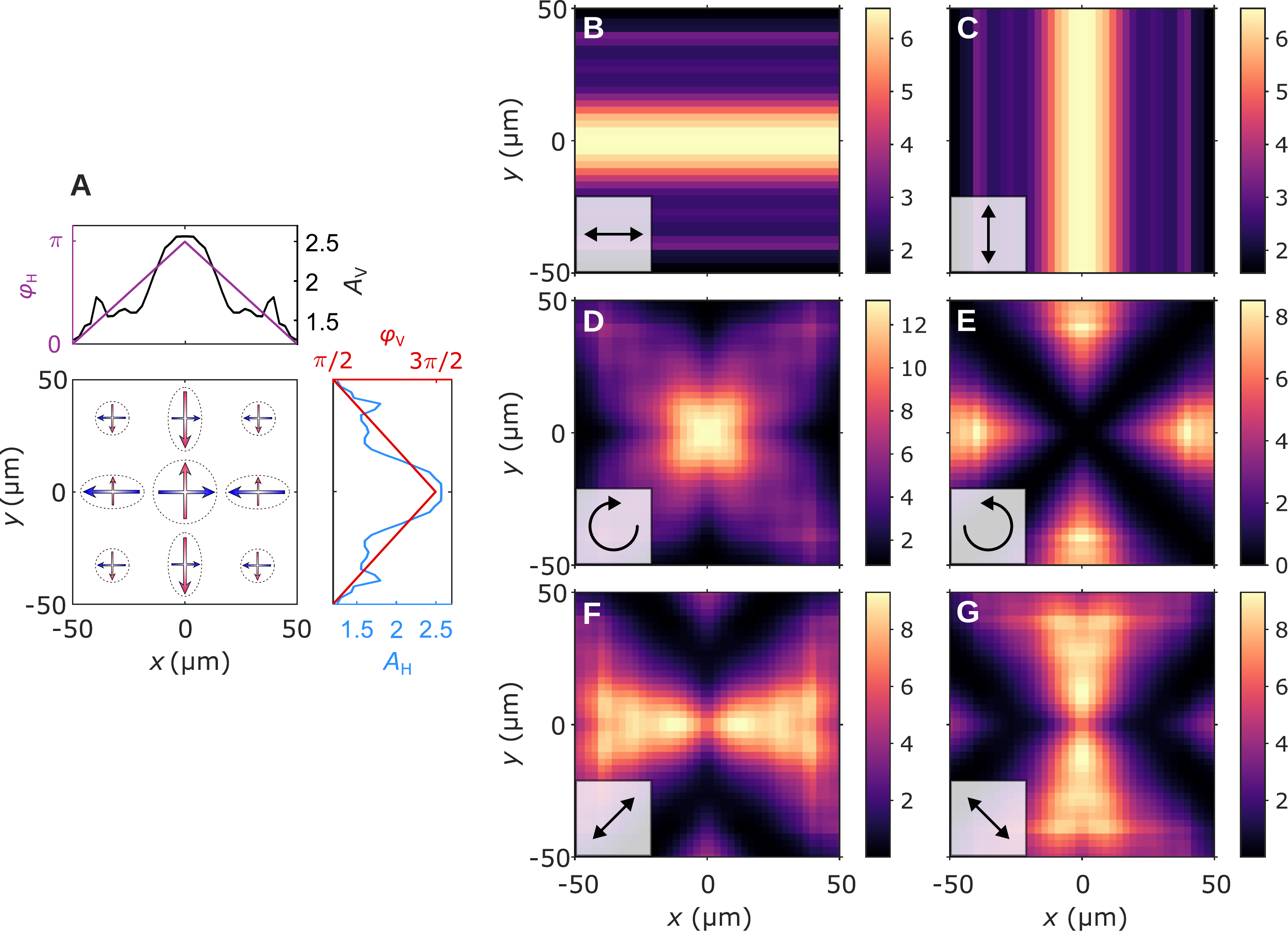}
\caption{\textbf{Jones vector model with fitted amplitude profiles.} (\textbf{A}) Modified version of the theoretical model shown in Fig.~4~(\textbf{A}). The linearly approximated amplitude profiles are replaced with averaged amplitude values from Fig.~2~(\textbf{B}). (\textbf{B}-\textbf{G}) Electric field intensities obtained by the Jones vector model presented in (\textbf{A}) with (\textbf{B}) horizontal, (\textbf{C}) vertical, (\textbf{D}) right circular, (\textbf{E}) left circular, (\textbf{F}) diagonal and (\textbf{G}) antidiagonal polarizers.}
\label{fig:figure_S3_fitted_model}
\end{figure}

The condensate intensity profiles are non-uniform as shown in Fig.~1~(\textbf{D}) without polarizers, and in Fig.~1~(\textbf{E}) and Figs.~2~(\textbf{A}-\textbf{F}) with polarization resolved. To illustrate this more clearly, we take the intensity values between $x = $ 0...50 \SI{}{\micro \meter} in Fig.~2~(\textbf{B}) and average them over the array in the $y$-direction: such profiles are shown in Fig.~\ref{fig:figure_S3_fitted_model}~(\textbf{A}). The origin of the non-uniform condensate intensity is in the finite size of the lattice, but the profiles are not linear. This is plausible, as the condensation process involves non-linearities (stimulated processes and saturation-induced effective photon-photon interactions~\cite{Vakevainen2020}). One may utilize the experimentally obtained intensity profiles in the theoretical model instead of the linear approximation for the decrease of electric field components; Figs.~\ref{fig:figure_S3_fitted_model}~(\textbf{B}-\textbf{G}) show that it leads to an even closer match between the experiment and the model. However, in the main text we chose to utilize the linear approximation in order to highlight the predictive power of the simple model.

\pagebreak

\pagebreak

\renewcommand\refname{References and Notes:}
\bibliography{science_arxiv}
\bibliographystyle{science_mod}

\end{document}